\documentclass{aa501}
\usepackage{graphicx}
\usepackage{natbib}
\newcommand{\eqb}{\begin{eqnarray}}
\newcommand{\eqe}{\end{eqnarray}}
\newcommand{\diff}{{\rm d}}

\newcommand{\obsdir}{\vec{\hat{n}}}
\newcommand{\rsheet}{r_{\rm s}}
\newcommand{\rlight}{r_{\rm L}}
\newcommand{\nlight}{n_{\rm L}}
\newcommand{\blight}{B_{\rm L}}
\newcommand{\rstart}{r_0}
\newcommand{\degree}{{\rm o}}
\newcommand{\sigmaT}{\sigma_{\rm T}}
\newcommand{\Pdot}{\dot{P}}
\newcommand{\nc}{\newcommand}
\nc{\nhotp}{n'_h}
\nc{\nhotpL}{n'_{h,L}}

\begin{document}
\title{Pulsed radiation from neutron star winds}
\author{J. G. Kirk\inst{1} 
 \and O. Skj\ae raasen\inst{1}
 \and Y. A. Gallant\inst{2}
}
\institute{Max-Planck-Institut f\"ur Kernphysik,
Postfach 10 39 80, 69029 Heidelberg, Germany
\and
Service d'Astrophysique, CEA-Saclay, F--91191 Gif-sur-Yvette, France}
\offprints{john.kirk@mpi-hd.mpg.de}
\date{Received \dots}
\abstract{
The radiation of a pulsar wind is computed
assuming that at roughly 10 to 100 light cylinder radii 
from the star, magnetic energy is dissipated into
particle energy. The synchrotron emission of heated
particles appears periodic, with, in general, both a pulse and an
interpulse. The predicted spacing agrees 
well with the Crab and Vela pulse profiles.
Using parameters appropriate for 
the Crab pulsar (magnetisation parameter at the light cylinder 
$\sigma_{\rm L}=6\times10^4$, Lorentz factor
$\Gamma=250$) agreement is found with the 
observed total pulsed luminosity. 
This suggests that the high-energy pulses
from young pulsars originate not in the corotating magnetosphere 
within the light cylinder (as in all other models) but from the
radially directed wind well outside it. 
}
\maketitle
\section{Introduction}
Recently, the non-axisymmetric 
\lq\lq striped pulsar wind\rq\rq\ investigated by
\citet{coroniti90} and \citet{michel94} has been reexamined
\citep{kirklyubarsky01}. A striped wind is a structure similar to a
Parker spiral, produced by the outward radial 
advection of magnetic field lines with foot points 
anchored on the surface of the rotating neutron star.
If the field lines originate from the poles of an 
obliquely rotating dipole, there exists 
a region around the equatorial plane where
the polarity of the field at a fixed radius reverses at
the rotation period, producing stripes of alternating
magnetic field direction.
Magnetic energy is dissipated into particle energy if reconnection occurs
at the stripe boundaries. \citet{coroniti90} 
proposed that this kind of
dissipation proceeds at a rate sufficient to maintain the thickness of the
current sheet approximately equal to the gyro-radius of the heated particles.
\citet{michel94} used essentially the same criterion, phrased in terms of the
velocity of the current carriers. These early papers concluded that 
observations of the Crab Nebula, which indicate conversion of Poynting flux
to particle-borne energy flux within $10^9\rlight$, 
(where $\rlight$ is the radius of the light cylinder)
could be explained by this
mechanism. However, \citet{lyubarskykirk01} point out that reconnection is
accompanied by acceleration of the wind. 
Both processes occur on the
same timescale in the Coroniti/Michel approach,
which is also the timescale on which the plasma
expands.
As a result, time dilation
in the accelerated flow significantly reduces the 
dissipation rate. In the case of the Crab,
conversion of Poynting flux to particle-borne energy flux is not achieved until
$10^{12}\rlight$ --- too slow to account for the
observations
[the $\sigma$-problem; see \citet{melatos98}].

However, our present
understanding of the physics of the dissipation  
process does not
exclude the possibility that it proceeds much more rapidly than
suggested by Coroniti and Michel. Recent investigations of 
reconnection in a relativistic pair plasma indicate that the 
timescale is related to the crossing time of a fast magnetosonic wave
\citep{zenitanihoshino01}. In the case of a transsonic wind, the process could
then be completed within a few pulsar periods, 
giving the plasma insufficient
time to accelerate to high Lorentz factor.
If rapid reconnection in fact occurs, it will be accompanied
by radiation losses \citep{melatos98} that may give rise to an observable
signature. 
In this letter we adopt this hypothesis.
We assume
that dissipation is 
triggered at a surface $r=\rstart$
well outside the light cylinder 
in the supersonic MHD wind of a pulsar
and takes place rapidly.  
Using two simple models of the electron distribution we 
show (i) that for the winds of young
pulsars
(e.g., the Crab) the predicted emission is pulsed, with, in general, a pulse
and interpulse that are not symmetrically spaced in
phase and (ii) that the predicted synchrotron 
luminosity in the pulses agrees to
order of magnitude with that observed in the high energy emission of
the Crab pulsar. 

Our calculations suggest that the 
high energy pulses emitted by young pulsars 
originate as synchrotron emission in the pulsar wind, far outside the light
cylinder. 
This is in marked contrast with conventional models of 
high-energy pulse production
at an inner or outer gap [for a review see
\citet{harding01}] and also with magnetospheric synchrotron based models
\citep{machabelietal00,crusiuswaetzeletal01}. The 
current sheet just beyond the region of closed field lines in the magnetosphere 
has 
been proposed as the source of high energy radiation from the 
Crab pulsar \citep{lyubarsky96}. But, in common with the 
others, this model relies for pulse production 
on the corotation with the star of a beam of radiation
emitted from within the light cylinder.

\section{Geometry of the emission region}
\label{geometry}

Any radial, relativistic flow containing a periodic modulation of the
emissivity is likely to appear pulsed to an observer because of two 
effects.
Firstly, the strong
Doppler boosting of approaching parts of the flow means that only a small
cone of the flow is visible, propagating within an angle of roughly $1/\Gamma$
with respect to the line of sight, 
where $\Gamma=(1-\beta^2)^{-1/2}$ is the Lorentz factor and 
$c\beta$ the radial
speed. The
spread in arrival times at the observer of photons emitted at the same radius 
but from different parts within this cone is roughly $r/(c\Gamma^2)$. Secondly,
if emission is restricted to a range $\Delta r$ in radius, the fact that the
photons move only slightly faster than the flow when seen in the laboratory
frame means that the spread in arrival times is roughly $\Delta r/(c\Gamma^2)$.
Thus, for a wind modulated at the period $P=2\pi\rlight/c$ of a rotating star, 
the approximate 
condition for the observation of pulsed emission is
\eqb
\Delta r,\ r&\la&\Gamma^2\rlight
\label{approximatepulse}
\eqe
\begin{figure}
\resizebox{\hsize}{!}
{\includegraphics[bb=74 278 705 507]{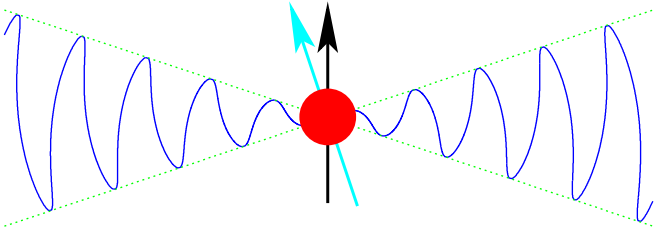}}
\caption{\protect\label{oblique}
A meridional section of the current sheet of an oblique
rotator
with a radial wind.
The thin lines limiting the extent of the sheet in latitude 
depict the plane of the magnetic equator when the magnetic axis
lies in the plane of the section.}
\end{figure}
If this is fulfilled, the observed pulse shapes appear similar to the 
pattern of modulation of the flow. In particular, for a flow containing
the corrugated spiral current sheet of an oblique rotator
(Fig.~\ref{oblique}), there should in
general be two components (pulse and interpulse) which are symmetrically placed
if the pulsar is viewed from the equatorial plane, but are asymmetrical when
viewed from higher latitudes. 
Denoting by $\alpha$ the angle between the
rotation axis and the magnetic axis, and by $\zeta$ the angle between the
rotation axis and the viewing direction, these pulses appear only for 
$|90^\degree -\alpha| < \zeta < 180^\degree-|90^\degree - \alpha|$. 

To improve on the estimate (\ref{approximatepulse}) and 
calculate the expected pulse shapes we adopt 
the following 
model of the volume emissivity $\epsilon$ of the wind:
\begin{itemize}
\item[(i)]
The emission region is located on the current sheet in the 
supersonic MHD wind
of a rotating neutron star with an oblique, split-monopole magnetic
field. 
\item[(ii)]
The dependence on frequency $\nu$ of the volume emissivity is a power law: 
$\epsilon\propto\nu^{-a}$.
\item[(iii)]
The emission switches on abruptly as plasma in the sheet crosses the surface
$r=\rstart$. Thereafter the volume emissivity decays as 
a power law in radius: $\epsilon\propto r^{-2-q}$.
\end{itemize}

Outside the light cylinder, assumption (i) 
should be a good approximation,
independent of the actual configuration of the magnetic
field close to the pulsar surface.  
The assumption that emission occurs only at the
exact position of the current sheet is an idealisation --- in reality, the entire flow may be
modulated, making the observed pulse more complex. 
In this case, our computations provide the Green's
function with which the pattern is to be folded to find 
the observed pulse shape.
If the wind carries enough particles to ensure MHD behaviour at least
initially [for an alternative approach see \citet{michelli99}], 
a simple
solution is available in the ultra-relativistic limit
\citep{bogovalov99}.
In it, the 
velocity is radial and constant and the magnetic field is almost toroidal.
The thin current sheet in such a wind is described by the equation
\eqb
r&=&
\rsheet(\theta,\phi,t)
\nonumber\\
&=&\beta\rlight\left[
\pm\arccos\left(-\cot\alpha\cot\theta\right)+c t/\rlight -\phi +2n\pi
\right]
\label{sheeteq}
\eqe
\citep{kirklyubarsky01}, where ($r$,$\theta$,$\phi$) are polar coordinates,
$t$ is the observer-frame time and $n$ is an integer.
 
Assumption (ii) ensures the
pulse shape is independent of frequency. In reality, the electron distribution
may exhibit breaks and cut-offs, which translates into different pulse
shapes at different frequencies. Our assumption, however, ensures that all
parts of the flow can contribute to the emission at a given frequency,
making it harder to obtain sharply defined pulses. 

\begin{figure}
\resizebox{\hsize}{!}
{\includegraphics[bb=40 159 566 689]{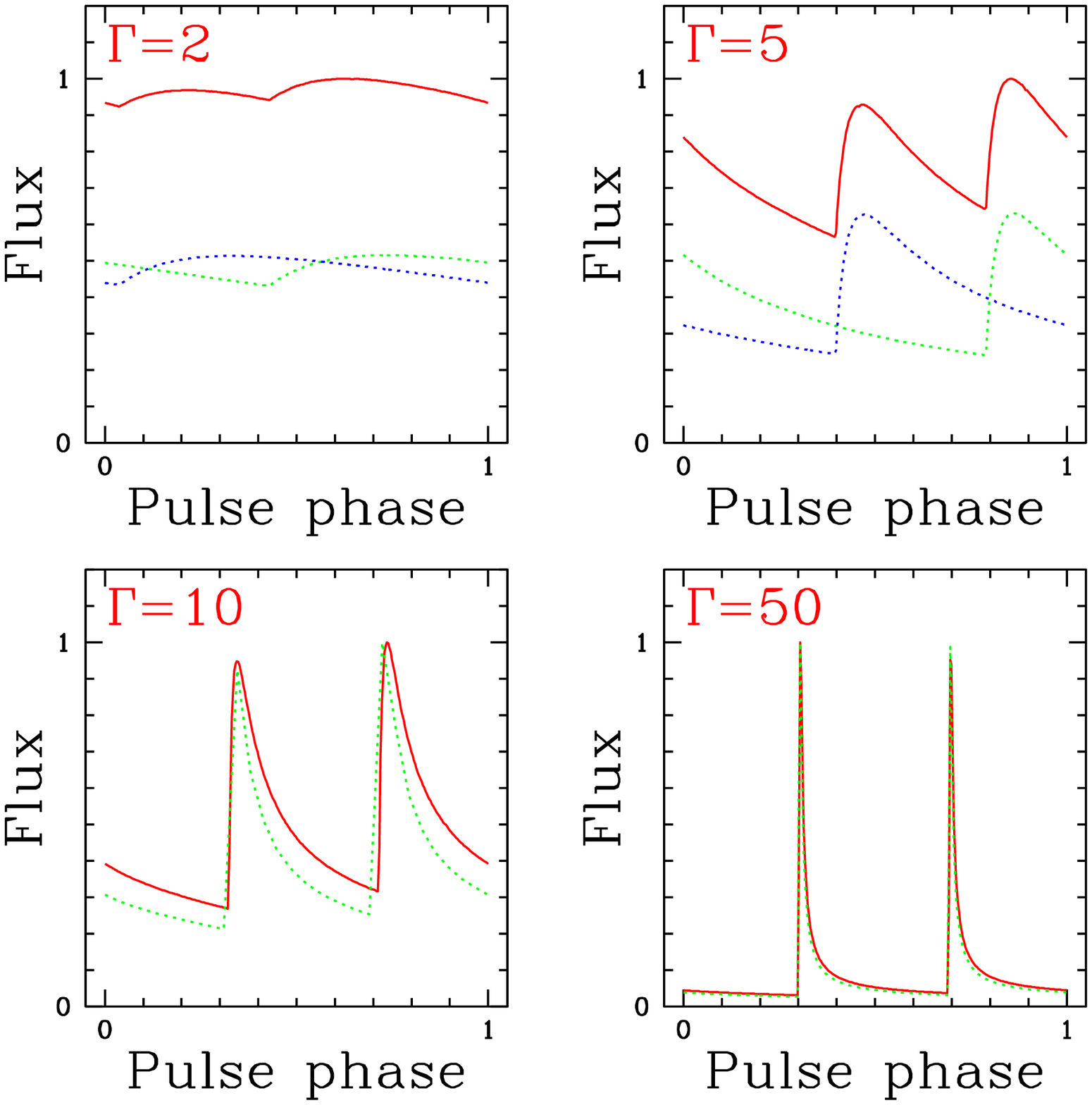}}
\caption{\protect\label{pulsefig}
Pulse shapes obtained by a numerical integration of Eq.~(\protect\ref{fluxeq}).
Four values of the Lorentz factor ($\Gamma$) are shown, 
the remaining parameters are: 
$\alpha=60^\degree$, $\zeta=60^\degree$, $\rstart=30\rlight$, $a=0$
and $q=3$ (see text). The total emission is shown as a solid line.
For $\Gamma=2$ and $5$, the contributions of the current sheets of different polarity
are shown as dotted lines.
For $\Gamma=10$ and $50$, the dotted line shows 
the effect of a more rapid fading of the
emissivity to larger radius ($q=6$).}
\end{figure}

The dependence of the observed flux
$F$ on $t$, $\alpha$ and the direction of observation $\obsdir$ is:
\eqb
F&\propto&
\int_{-\infty}^{+\infty}\diff t'\int_{\rstart}^\infty \diff r
\int_{90^\degree-\alpha}^{90^\degree+\alpha} \diff\theta\int_0^{2\pi}\diff\phi
\nonumber\\
&&r^{-q}D^{2+a}\delta\left[r-\rsheet(\theta,\phi,t')\right]
\delta\left(t'-t+\obsdir\cdot\vec{r}/c\right)
\label{fluxeq}
\eqe
where $D=1/\left(1-\beta\vec{\hat{r}}\cdot\obsdir\right)$ is the 
conventionally defined Doppler 
factor, apart from a (constant) factor of $\Gamma$.
Choosing $a=0$ and $q=3$, which is appropriate
for the synchrotron radiation of a fixed number of particles that undergo
predominantly adiabatic losses in the expanding plasma,
we present sample pulses for various Lorentz factors in
Fig.~\ref{pulsefig}. For the higher Lorentz factors this figure also
displays the effect of a faster fall-off of the emissivity towards
higher radii ($q=6$), mimicking the effect of an accelerating wind. The
asymmetry in time of the pulses is produced by our assumption of a
sudden switching on 
of the emission and is reduced by a more rapid switching off.
The parameters of this figure have been chosen to correspond to the Crab
pulsar. 
The radio emission of this object is complex.
The shape of the high energy pulses is repeated in the radio, suggesting a
common emission site, which in our model would be the wind. But there are
additional radio components \citep{moffatthankins96}, one of which displays
\lq core emission\rq\ properties \citep{rankin90}.
This suggests emission from close to the polar cap, with the magnetic
axis passing within a few degrees of the line of sight. The core
emission is in the precursor of the main pulse, which constrains 
the relative phase of the polar cap emission and wind emission 
in this interpretation. However, a 
complete solution linking the inner magnetosphere and the wind structure 
would be required to check this point. Nevertheless, core emission
indicates that the 
inclination angle $\alpha$ of the magnetic axis to the rotation axis
roughly equals the angle $\zeta$ 
between 
the rotation axis and the line of sight. The latter  
is determined by X-ray and
optical observations of the \lq torus\rq\ 
\citep{aschenbachbrinkmann75,hesteretal95} to be $60^\degree$. 
In Fig.~\ref{pulsefig}
the emission has been assumed to switch on at $\rstart=30\rlight$ 
and pulses appear for $\Gamma>5$, in rough agreement with 
the estimate (\ref{approximatepulse}). At higher Lorentz factors, the 
pulses sharpen. From Eq.~(\ref{sheeteq}) it can be seen that the
emission splits naturally into two contributions, corresponding to 
different signs of the term $\arccos(-\cot\alpha\cot\theta)$. In
the first two plots of Fig.~\ref{pulsefig} these contributions are plotted
separately as dotted lines. 
Although the detailed pulse shapes will depend on the
structure of the reconnecting sheets in the Crab wind, as well as the 
details of the switching on mechanism and the amount of acceleration
of the wind, the separation 
of the two peaks is not sensitive to these unknowns, 
being determined by the location of the
sheets. This separation in phase equals 
$\arccos(\cot\alpha\cot\zeta)/180^\degree$,
corresponding to $0.39$ for the Crab pulsar, 
in good agreement with observation [e.g., \citet{kanbach98,kuiperetal01}].

For the Vela pulsar, conflicting interpretations of the 
polarisation sweep and the
X-ray morphology give $\alpha\approx71^\degree$,
$\zeta\approx65^\degree$ \citep{radhakrishnandeshpande01} 
or
$\alpha\approx\zeta\approx53^\degree$ 
\citep{helfandetal01}. These imply separations of the
gamma-ray (sub-)pulses of $0.45$ and $0.31$ of a period respectively, 
so that the observed 
spacing of $0.4$ marginally favours the former interpretation.

\section{Luminosity estimates}
\label{luminosity}
Electrons heated in the current sheets will enter the
surrounding magnetic field and emit synchrotron radiation. An estimate
of the power radiated 
can be found by assuming a monoenergetic electron population 
of Lorentz factor $\gamma$ ($\gg1$) and
requiring pressure
equilibrium between the hot plasma and the surrounding magnetic field. 
In the delta-function approximation for the synchrotron emissivity, 
the luminosity at frequency $\nu$ into 
a solid angle $\Omega$ is
\eqb
{\diff L/\diff\nu\diff\Omega}&\approx&\int_{\rstart}^\infty\diff r\,
r^2 \Delta\Gamma n'_{\rm h}P_{\rm s.p.}\delta\left(\nu-\nu_0\right)
\label{logos}
\eqe
\citep{skjaeraasenkirk01}, where $\Delta$ ($\le1$) is the fraction of a
wavelength of the wind pattern occupied by hot electrons of proper density
$n'_{\rm h}$, $P_{\rm s.p.}=(4/3)c\sigmaT\gamma^2(B'^2/8\pi)$
is the synchrotron power emitted by a 
single electron, and $\nu_0$ is the characteristic frequency of emission, 
measured in the observer frame ($\sigmaT$ is the Thomson cross section). 
Pressure equilibrium implies
$n'_{\rm h}\gamma mc^2/3= B'^2/8\pi$, where $B'$ is the magnetic field in
the comoving frame and $m$ the electron mass. 
Well outside the light cylinder, but 
before energy release begins, i.e., at $\rlight<r<\rstart$, 
the density and magnetic field scale as $r^{-2}$ and $r^{-1}$ respectively, 
and can be characterised by the values $\nlight$ and $\blight$ 
obtained by extrapolating 
back to $r=\rlight$. Provided the flow does not accelerate significantly 
whilst emitting synchrotron radiation, these same scalings,
which imply $\gamma$ independent of $r$, can be used 
in the integration in Eq.~(\ref{logos}). 
This is convenient for an estimate of the luminosity, but is sensitive to
the assumption that the wind does not accelerate upon reconnection.

If we identify the ratio of Poynting flux to kinetic energy (at the
light cylinder)
$\sigma_{\rm L}=\blight^2/(4\pi\Gamma \nlight mc^2)$, 
define an average fraction of hot particles
$\bar{\Delta}=\rstart\int_{\rstart}^\infty\diff r\Delta
/r^2\la1$, and use the standard relation
$\blight=2.9\times10^8\Pdot^{1/2}P^{-5/2}\,$G, the luminosity becomes
\eqb
{\diff\hat L\over\diff \Omega}
&=&
3.9 \times10^7\left({\sigma
\bar{\Delta}\rlight\Pdot\over\Gamma^3\rstart P^4}\right)
\ (P\ {\rm in\ secs})
\label{finalflux}
\eqe
where $\diff\hat L/\diff \Omega$
is the synchrotron luminosity per steradian divided 
by the wind luminosity per steradian.
Similarly, the characteristic frequency is
\eqb
h\nu_0(r)&=&
2.6\times10^{-6}\left({\rlight/r}\right)\sigma^2\Pdot^{1/2}P^{-5/2}
\ {\rm MeV}.
\label{freq1}
\eqe

The spectrum of this model with
monoenergetic electrons extends to a maximum frequency
given by Eq.~(\ref{freq1}) with $r=\rstart$ and 
is flat:
$\diff L/\diff\nu\diff\Omega\propto \nu^0$.  However, this aspect is 
sensitive to expansion and acceleration of the 
plasma after 
reconnection.
If, in a more realistic 
scenario, the electron density is a power law 
$N(\gamma)\propto\gamma^{-p}$, with $p>2$, then most of the energy resides in 
the particles with lowest $\gamma$ and the above estimate (\ref{freq1}) 
refers to the photons emitted by these particles. The 
emission at frequencies $\nu<\nu_0(\rstart)$ 
is dominated by the same low energy particles
radiating at larger radius
in a weaker magnetic field, and is determined by the expansion and 
acceleration of the hot plasma. In the absence of acceleration, the flat
spectrum derived above
prevails.
However, the intrinsic spectrum, which is a power law 
$\diff L/\diff\nu\diff\Omega\propto \nu^{-a}$,
with $a>0.5$, is 
revealed at frequencies $\nu>\nu_0(\rstart)$, where particles
radiating at $\rstart$ dominate.
In the case of the Crab pulsar, 
the optical to X-ray emission is consistent with a flat spectrum
\citep{shearergolden01}, which steepens to $a>1$ 
at a photon energy of 
approximately $1\,$MeV \citep{kanbach98,kuiperetal01}. Equation~(\ref{freq1})
then
implies
$\sigma=1.1\times10^4\sqrt{\rstart/\rlight}$. Observations suggest
$\diff\hat L/\diff\Omega\approx 10^{-3}$, which,
from Eq.~(\ref{finalflux}), is consistent with a mildly supersonic flow,
$\Gamma\approx 250$, that starts to radiate at $\rstart=30\rlight$. 
This satisfies the lower limit of $5\rlight$ \citep{bogovalovaharonian00} 
on the radius at which Poynting flux 
can be converted to particle-borne energy flux.

\section{Discussion}
\label{discussion}
Current models of the high energy emission from rotation powered
pulsars fall into two groups: \lq\lq polar cap\rq\rq\ and 
\lq\lq outer gap\rq\rq\ models [see 
\cite{harding01}]. Each locates the emission region in the
corotating magnetosphere of the star. The predicted pulse shapes 
depend sensitively on the uncertain geometry of the magnetic field in
this region, usually assumed dipolar. 
In contrast, in our model, pulses are emitted outside the corotation
region. If the pulsar drives a supersonic, MHD wind, the
field geometry there approaches a simple asymptotic solution
\citep{bogovalov99}, determining the basic properties of the
pulses. 

\citet{pacini71} and \citet{pacinisalvati83,pacinisalvati87} 
suggested that particles close to the light cylinder emit 
synchrotron radiation and are responsible for the optical pulses of the
Crab and other pulsars, a theory which appears to be in reasonable
agreement with observation \citep{shearergolden01}.
Our model is similar in that the radiation
mechanism is synchrotron emission and the scaling of Eq.~(\ref{finalflux}) 
is close to that originally given by \citet{pacini71}. It goes further,
however, by specifying for the emission region
a precise geometrical structure and location.

It is known that the dilution of the plasma 
in the wind must lead to non-ideal MHD behaviour  \citep{usov75,michel82}
and so can be responsible for triggering the emission. 
The position at which this occurs depends on the 
initial concentration of charges in the current sheet and is taken as
a free parameter. The major uncertainty in the model is the speed with
which dissipation proceeds. Our computations implicitly assume rapid
dissipation (over a scale small compared to the radius). In this
case, it is possible to make a rough estimate of the synchrotron
luminosity which is in agreement with observations. 
It is conceivable that
at least part of the radio emission could be produced in the wind region, 
as is suggested by the similarity of the pulse profiles 
at all frequencies seen in the Crab
pulsar. However, until a candidate coherent mechanism can be 
identified, this connection remains very speculative.
Nevertheless, the wind scenario is 
a viable alternative to current theories 
of gamma-ray, X-ray and
optical pulses from rotation-driven pulsars and is capable of making 
testable predictions about the pulse profile. 

\acknowledgement{This work is a collaboration of the TMR Network 
\lq\lq Astroplasmaphysics\rq\rq\ of the European Commission, 
contract FMRX-CT-98-0168. Y.G. is supported by a Marie Curie Fellowship 
from the European Community, contract no. HPMFCT-2000-00671 under the 
IHP programme.}

\end{document}